\documentclass[superscriptaddress,twocolumn,prx]{revtex4}

\usepackage{graphicx}% Include figure files
\usepackage{subfigure}
\usepackage{float}
\usepackage{bm}% bold math
\usepackage{hyperref}% add hypertext capabilities
\usepackage{amsmath}
\usepackage{amssymb}
\usepackage{amsthm}
\usepackage[normalem]{ulem}
\begin{document}

%---------------------------------------------------------------------------------------------------------
\title{The formation of the positive, fixed charge at c\mbox{-}Si(111)/a\mbox{-}Si$_3$N$_{3.5}$:H interfaces}
%---------------------------------------------------------------------------------------------------------

\author{L.~E.~Hintzsche}
\affiliation{University of Vienna, Faculty of Physics and Center for 
Computational Materials Science, Sensengasse 8/12, A-1090 Vienna, Austria}

\author{C.~M.~Fang}
\affiliation{University of Vienna, Faculty of Physics and Center for 
Computational Materials Science, Sensengasse 8/12, A-1090 Vienna, Austria}

\author{M.~Marsman}
\affiliation{University of Vienna, Faculty of Physics and Center for 
Computational Materials Science, Sensengasse 8/12, A-1090 Vienna, Austria}

\author{M.~W.~P.~E. Lamers}
\affiliation{ECN Solar Energy, P.O. Box 1, 1755 ZG Petten, The Netherlands}

\author{A.~W.~Weeber}
\affiliation{ECN Solar Energy, P.O. Box 1, 1755 ZG Petten, The Netherlands}

\author{G.~Kresse}
\affiliation{University of Vienna, Faculty of Physics and Center for 
Computational Materials Science,
 Sensengasse 8/12, A-1090 Vienna, Austria}

%---------------------------------------------------------------------------------------------------------
%   abstract
%---------------------------------------------------------------------------------------------------------

%---------------------------------------------------------------------------------------------------------
\begin{abstract} 
Modern electronic devices are unthinkable without the well-controlled formation
of interfaces at heterostructures. These often involve at least one amorphous
material. Modeling such interfaces poses a significant challenge, since a
meaningful result can only be expected by using huge models or by drawing from
many statistically independent samples. Here we report on the results of high
throughput calculations for interfaces between crystalline silicon
(c\mbox{-}Si) and amorphous silicon nitride (a\mbox{-}Si$_3$N$_{3.5}$:H), which
are omnipresent in commercially available solar cells. The findings reconcile
only partly understood key features. At the interface, threefold coordinated Si
atoms are present. These are caused by the structural mismatch between the
amorphous and crystalline part. The local Fermi level of undoped c\mbox{-}Si
lies well below that of a\mbox{-}SiN:H. To align the Fermi levels in the
device, charge is transferred from the a\mbox{-}SiN:H part to the c\mbox{-}Si
part resulting in an abundance of positively charged, threefold coordinated Si
atoms at the interface. This explains the existence of a positive, fixed charge
at the interface that repels holes. 
\end{abstract}
%---------------------------------------------------------------------------------------------------------

\maketitle

\section{Motivation}

% Importance of c-Si/a-SiN interfaces

Silicon is the most important material for single-junction solar cells.
Although there are other promising compounds for photo voltaic devices such as
GaAs, CdTe, or InP, silicon based cells still maintain about $90$~\% market
share \cite{Luque2011,AbouRas2011}. Silicon has the advantage of being widely
available, non-toxic, and applicable  to many different requirements. The
possible applications range from less efficient, but cheap, amorphous cells
over cost-efficient multi-crystalline cells up to highly efficient, but more
expensive, mono-crystalline cells \cite{Luque2011,Green2012}. To produce high
performance cells, optical losses must be minimized and recombination centers
passivated. Meeting both challenges, amorphous silicon nitride (a\mbox{-}SiN:H)
is commonly deposited as anti-reflection and passivation layer on top of
crystalline silicon (c\mbox{-}Si) solar cells with p-type base and n-type
emitter. Using plasma enhanced chemical vapor deposition (PECVD)
\cite{Aberle2000,Duerinckx2002}, the deposition parameters (i.e.  the gas
mixture of N$_2$/SiH$_3$ or NH$_3$/SiH$_3$) control the refractive index as
well as the defect concentration of a\mbox{-}Si$_3$N$_x$:H so that the
reflectivity and the passivation can be tuned for optimal cell performance.

% defect centers and experimental limitations

However, even with high quality surface passivation, the defect concentration
at interfaces is larger than in bulk materials, and defect assisted
Shockley-Read-Hall (SRH) recombination remains a major source of carrier losses
in solar cells \cite{Shockley1952,Hall1952}. Especially, defect levels in the
middle of the Si band gap are effective recombination centers and carrier traps.
They reduce the lifetime of electron-hole pairs in c\mbox{-}Si and, therewith,
the efficiency of the solar cells \cite{Bube1983,Green1986}. Lamers~{\em
et~al.} showed that the defect density correlates with what is called the
"fixed charge" \cite{Lamers2012}. This fixed charge is positive and located at
the c\mbox{-}Si/a\mbox{-}SiN:H interface. In contrast to charge neutral
recombination centers,  the fixed charge increases the lifetime of charge
carriers in n-type doped c-Si by repelling holes from the interface. As
undercoordinated Si atoms can have negative, neutral, or positive charge (i.e.
K$^-$, K$^0$, or K$^+$ defects), the fixed charge is typically explained by an
increased number of K$^+$ defects at the interface \cite{Aberle2001}.  However,
it is still unclear how those defects form and whether one could also produce
c-Si/a-SiN:H interfaces with a negative, fixed charge that repels electrons and
attracts holes. It is therefore obvious that defects are important to
understand the properties of c\mbox{-}Si/a\mbox{-}SiN:H interfaces.
Nonetheless, they are very difficult to investigate by experimental methods
alone. For instance, lifetime measurements estimate defect concentrations
indirectly, and electron spin resonance (ESR) measurements can solely detect
states occupied by a single electron \cite{Robertson1995}.

% Previous studies on c-Si/a-SiN:H interfaces

For this reason, several authors recently examined c\mbox{-}Si/a\mbox{-}SiN:H
interfaces by using computer
simulations \cite{Omeltchenko2000,Yang2009,Butler2011,Butler2012,Larsen2013,Lamers2013,Lamers2013-2,Pham2013}.
For example, Butler~{\em et~al.} performed a topological analysis and compared
the density of structural defects for different thicknesses of the transition
region between c\mbox{-}Si and a\mbox{-}SiN:H \cite{Butler2011,Butler2012}. The
study suggests an optimal thickness of $2$~nm, however, a computationally cheap
Tersoff potentials was used and, like other studies employing force field
methods \cite{Omeltchenko2000}, their work was limited to investigations of
structural properties. Specifically, their conclusions on electronic properties
were based on the assumption that all undercoordinated atoms are recombination
centers. As the electronic properties are important for most applications, {\em
ab initio} methods seem to be more suitable.  Pham~{\em et~al.} generated
c\mbox{-}Si/a\mbox{-}SiN:H interface structures by combining classical
molecular dynamics and ab intio methods \cite{Pham2013}. Afterwards, they
examined the G$_0$W$_0$ corrected band offsets and the change of the dielectric
constant perpendicular to the interface. Their focus clearly lied on the
application of a\mbox{-}SiN:H as high K dielectric in MOS-FET devices, since
recombination centers were not examined in their work.

\section{Modeling Setup}

% Our contribution to the scientific community

In the present paper, we explicitly investigate the electronic properties of
gap states at c\mbox{-}Si/a\mbox{-}Si$_3$N$_{3.5}$:H interfaces with $1$ and
$11$~atm\% hydrogen (H). The undoped c\mbox{-}Si part is modeled by a hexagonal
$\sqrt{7}\times\sqrt{7}$ unit cell containing $4$ double layers of $7$ Si atoms
stacked in the ($111$)-direction with a corresponding lattice constant of
$a=5.43$~\AA. The a\mbox{-}SiN:H part is made up by  $59$ Si, $69$ N and $1$ or
$17$ H atoms. We have also investigated $4 \times 4$ unit cells with $6$ Si
double layers of $16$ Si atoms finding virtually identical results.  The
respective densities and other computational parameters are  chosen in
accordance with our previous study \cite{Hintzsche2012}. Since defects are
minority species which are difficult to describe using force fields, our study
uses a large scale ab initio molecular dynamics approach based on density
functional theory (DFT)
\cite{Bloechel1994,Kresse1994,Kresse1996,Perdew2008,Hintzsche2012}.  $90$
statistically independent samples are prepared by cooling
a\mbox{-}Si$_3$N$_{3.5}$:H from the melt ($2500-2700$~K) to the amorphous state
($1500$~K) in $37.5$~ps. To compensate for the lack in system size, we
afterwards average over the microscopic configurations.  Finally, we examine
the electronic gap states and link them to the geometrical properties of the
structures \cite{Hintzsche2013}. Before we continue, we note that our
interfaces are sharp. This has been achieved by keeping the c\mbox{-}Si atoms
partially fixed by allowing only for in plane movements at the interface.
Eventually, we have removed this restriction at the interface in the final
relaxation step.

\section{Results}

\subsection{Spatial location of gap states}
% Localized and de-localized gap states

%-------------------------------------------------------------------------
\begin{figure}
    \begin{center}
        \includegraphics[trim= 0.15cm 0.5cm 0.2cm 0.0cm, width=8.5cm]{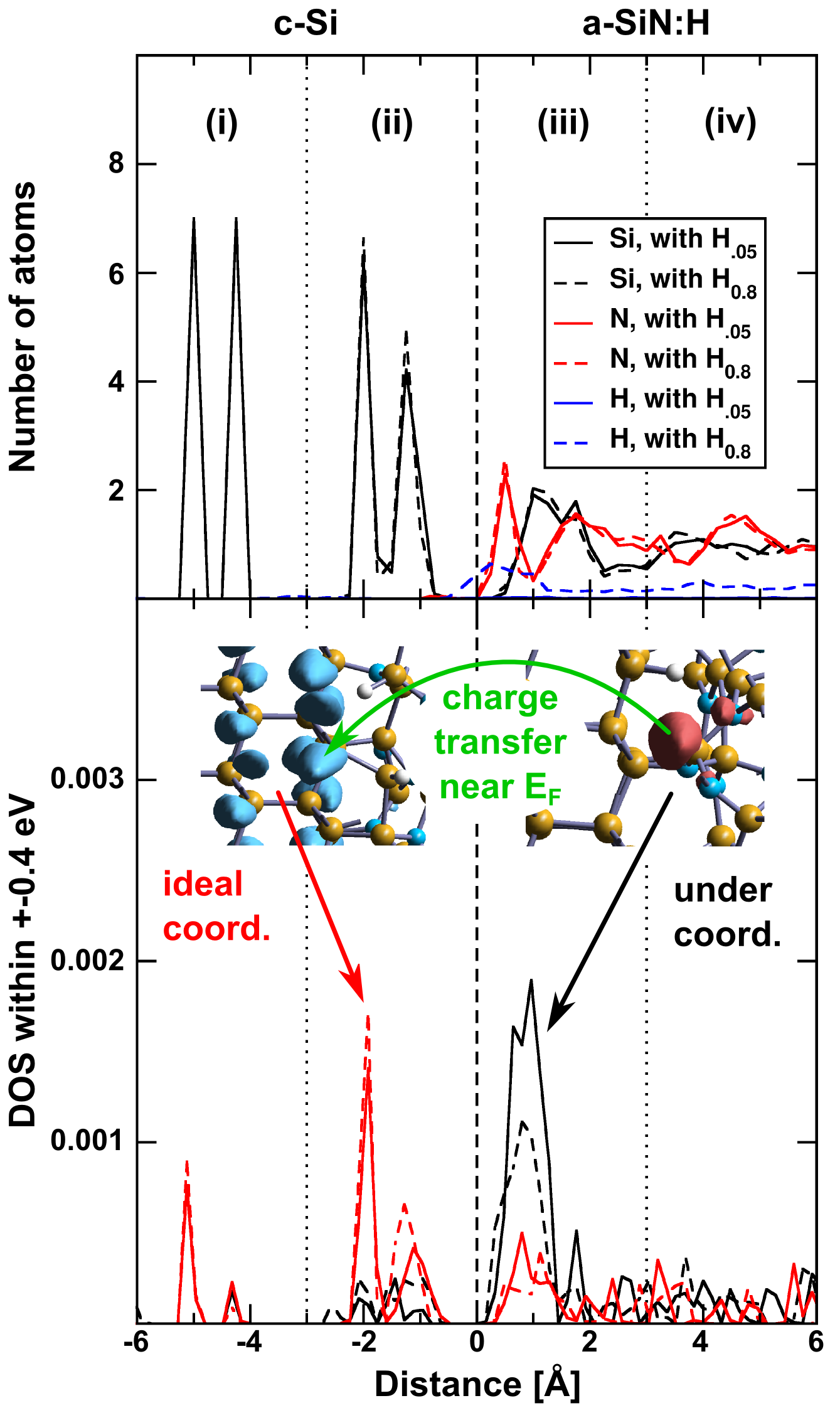}
    \end{center}
    \caption{
(color online) (a) Number densities and (b) layer resolved density of states
(DOS) at the c\mbox{-}Si/a\mbox{-}Si$_3$N$_{3.5}$:H interfaces with $1$ and
$11$~atm\% H (solid/dashed lines).  (a) The N number density shows a peak close
to the interface integrating to $60$~\% of the number density in the top most
Si layer. (b)  The considered electronic states lie within $\pm0.4$~eV of the
Fermi level. The states are found to be predominantly localized either at atoms
with ideal- or under-coordination (red/black arrow). These states are related
to  the conduction band of c\mbox{-}Si or to undercoordinated Si atoms in
a\mbox{-}Si$_3$N$_{3.5}$:H. In some microscopic models, a charge transfer from
occupied coordination defects (red isosurface) to the Si conduction band (blue
isosurface) occurs (green arrow). 
}
    \label{fig:NrDefectProfile}
\end{figure}
%-------------------------------------------------------------------------

In the following, we summarize our most important results by investigating the
layer resolved number densities and the electronic density of states (DOSs) at
the c\mbox{-}Si/a\mbox{-}Si$_3$N$_{3.5}$:H interface (Fig.~\ref{fig:NrDefectProfile}).  The
structural mismatch causes slight distortions in the first double layer of
c\mbox{-}Si, and more important, long range density fluctuations in
the a\mbox{-}SiN:H part, with a pronounced increase in the N density around
$1.7$~\AA\ (upper panel, red lines) above the Si topmost layer. These N atoms
always bind to Si surface atoms, saturating  $60$~\% of the Si surface dangling
bonds. The other Si surface atoms bind to Si atoms in the amorphous part. This
causes the onset of a peak in the Si number density. The corresponding peak
integrates to $1.3$ Si atoms per Si surface atom, and the region is unusually
Si rich, with very little N content. Analysis of the bonding topology suggests
that a large number of threefold coordinated Si K defects exists in this
region, similar to, but in much greater number than in bulk
a-SiN:H \cite{Hintzsche2013}. These geometrical defects also cause a large
number of electronic defect levels located in the band gap of c\mbox{-}Si
[lower panel, region (iii)]. On the c\mbox{-}Si side, our analysis at first
sight  also  suggests a large number of defects levels [lower panel, region
(ii)], however, inspection of the charge density of these states indicates that
these are conduction band like c\mbox{-}Si states (red arrow).  We note that
analysis of the large $4 \times 4$ unit cell yields {\em quantitatively}
identical results (not shown).

One  central finding is that charge is partially transfered from originally
occupied K defects in the a-SiN:H  to the c\mbox{-}Si conduction band.  This
happens in $7$~\% of our microscopic models (green arrow). This result, which
is substantiated in detail below, explains the existence of the positive, fixed
charge at c\mbox{-}Si/a\mbox{-}SiN:H interfaces. When H is present, the number
density of H is clearly increased in the a\mbox{-}SiN:H part at the interface
(upper panel, blue line), but the N and Si number densities are hardly changed.
Although H reduces the number of K defects at the interface almost by a factor
$2$ (lower panel, solid/dashed black line), our general observations remain
unchanged.

%----------------------------------------------------------------
\begin{figure}
\begin{minipage}{0.475\textwidth}
    \centering
    \includegraphics[trim= 0.0cm 0.0cm -0.1cm 0.0cm, width=8.5cm]{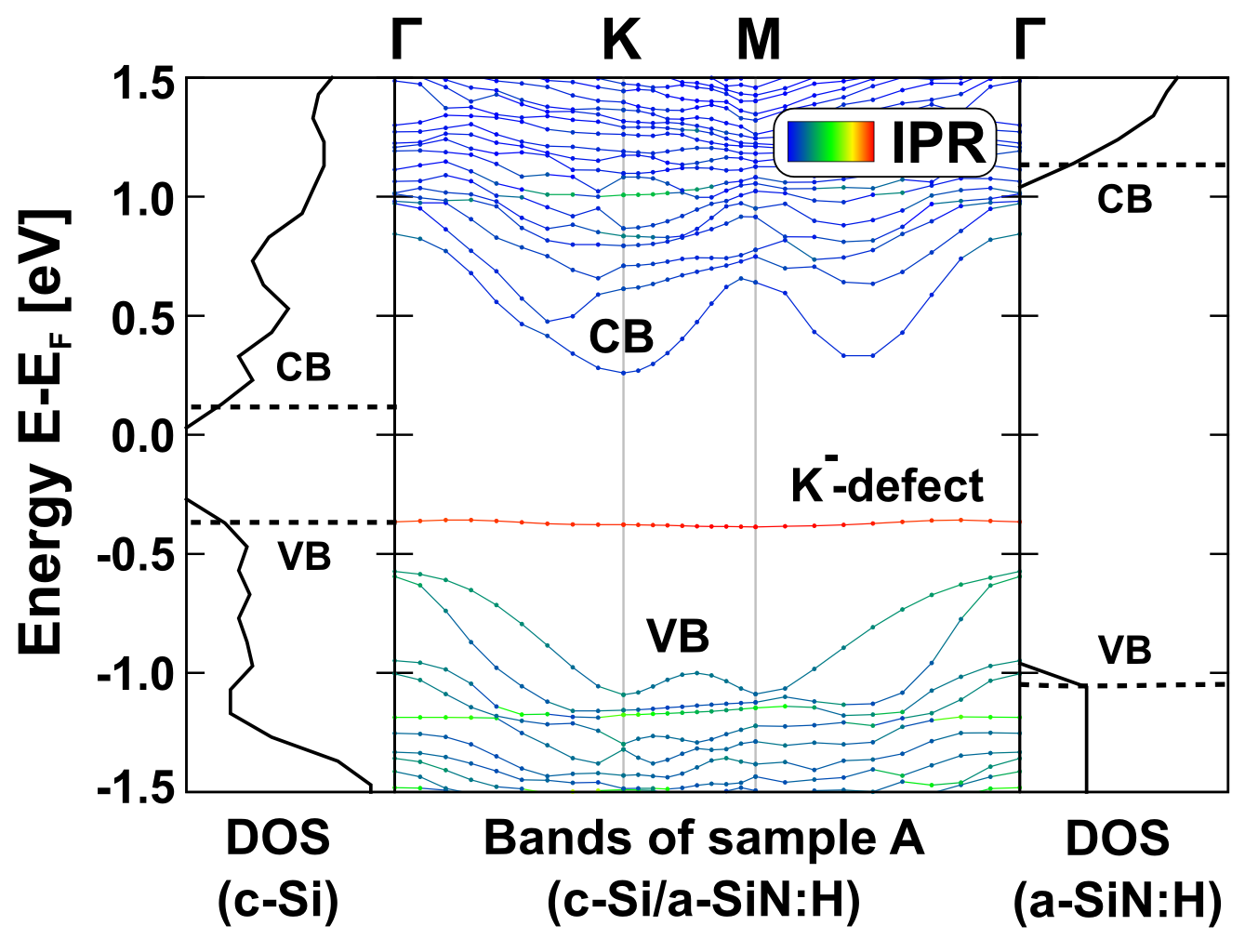}
    \caption{
(color online) Typical band structure and DOS of a
c\mbox{-}Si/a\mbox{-}Si$_3$N$_{3.5}$:H interface sample in the presence of a
single defect state. The color coding indicates the degree of localization and
corresponds to the inverse participation ratio (red for localized states).
Additionally, we show the DOS of a bulk undoped c\mbox{-}Si and a defect free
bulk a\mbox{-}Si$_3$N$_{3.5}$:H sample.  Note that E$_{\text{F}}$ refers to the
average Fermi level of the interface models.
    }
    \label{fig:bandsDosIpr}
\end{minipage}
\hfill
\begin{minipage}{0.475\textwidth}
    \centering
    \includegraphics[trim= 0.0cm 0.0cm 0.0cm 0.0cm, width=8.5cm]{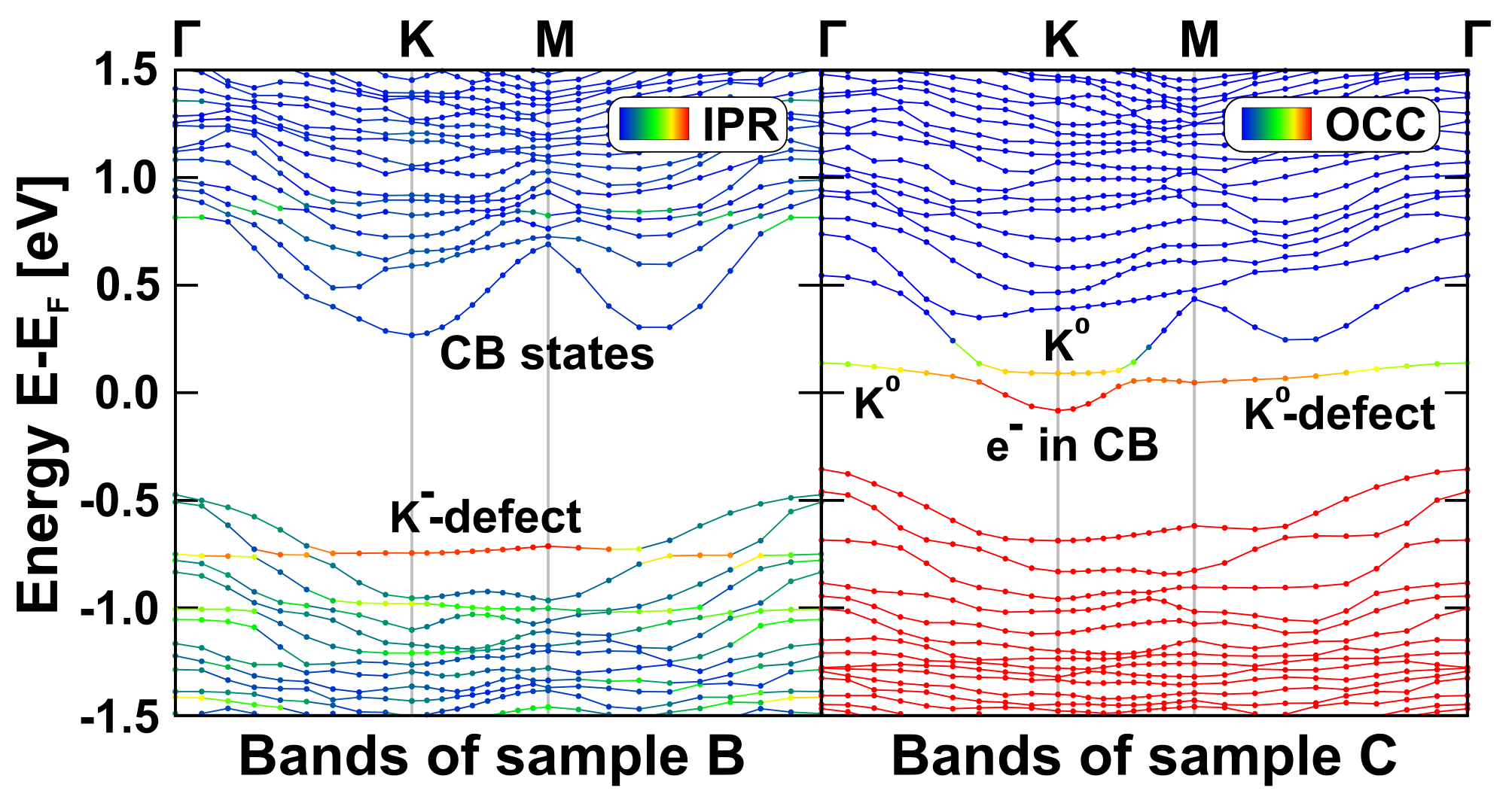}
    \caption{
(color online) Two examples for occupied K defects. In sample B, the defect
level is close to the valence band, whereas, in sample C, it is close to the
conduction band. In sample C, a partial charge transfer into the conduction
band is observed. Both defect states interact with the valence or conduction
band states resulting in forbidden crossings. 
}
    \label{fig:bandCrossing}
\end{minipage}
\end{figure}
%----------------------------------------------------------------

\subsection{Band structure and defect levels}

% Prototype band structure and defect level

We now investigate three exemplary microscopic models in more detail. The first
sample A contains a single localized doubly occupied K$^-$ defect in the middle
of the gap (Fig.~\ref{fig:bandsDosIpr}). In addition to the band structure of
sample A, we show the DOSs of bulk undoped c\mbox{-}Si and a defect free bulk
a\mbox{-}SiN:H model structure (left and right most panel). These bulk states
have been aligned at the $1s$ core levels of the Si and the N atoms in the
middle of the c\mbox{-}Si and a\mbox{-}SiN:H part, respectively.  The
comparison of the bulk DOSs  shows that the indirect band gap of c\mbox{-}Si
(about $0.5$~eV) is much smaller than the band gap of a\mbox{-}SiN:H (about
$2.3$~eV), and the conduction band offset (about $1.0$~eV) is larger than the
valence band offset (about $0.8$~eV). That the band gaps are too small compared
to experiment is a well known DFT artifact, however, as shown for instance in
Ref.  \onlinecite{Larsen2013}, this error changes the mid gap levels, which are
particularly relevant for this study, only very little. The present offsets are
consistent with photo emission experiments and previous simulations on
c\mbox{-}Si/c\mbox{-}Si$_3$N$_4$ interfaces \cite{Kaercher1984,Larsen2013}. The
band structure furthermore demonstrates that the in plane dispersion of the
defect level is negligible.  This indicates that the lateral supercell size is
generally sufficient, as there is no interaction between the periodically
repeated defect images.  However, the valence and the conduction band of
c\mbox{-}Si show a strong dispersion necessitating a dense $8\times 8$ k-point
sampling parallel to the interface.

% Overlap between band and defect states

Since defect related states in the amorphous part can be situated anywhere
within the gap of a\mbox{-}SiN:H \cite{Hintzsche2013}, they often overlap with
the c\mbox{-}Si conduction band or valence band. To give examples for such
situations we show the band structure of two samples B and C
(Fig.~\ref{fig:bandCrossing}).  In sample B, a doubly occupied K$^-$ defect
merges into the c\mbox{-}Si valence band, and, in sample C, an initially doubly
occupied K$^-$ defect merges into the c\mbox{-}Si conduction band. In both
samples, the defect states are again strongly localized and practically flat.
Nevertheless, since small hybridizations between the c\mbox{-}Si and the defect
level exist, forbidden crossings are observed. The most interesting case is
sample C. Here the average defect level is slightly above the conduction band
minimum of c\mbox{-}Si, causing a partial transfer of charge from the defect
level into the c\mbox{-}Si conduction band. This has reduced the occupancy
(OCC) of the defect from K$^-$ to K$^0$ and relates to the main finding
mentioned above.  The opposite, a partial transfer from the c\mbox{-}Si valence
band to an originally empty K$^+$ defect, is never observed.

%-------------------------------------------------------------------------
\begin{figure}
    \begin{center}
        \includegraphics[trim= 0.0cm 0.0cm 0.0cm 0.0cm, width=8.5cm]{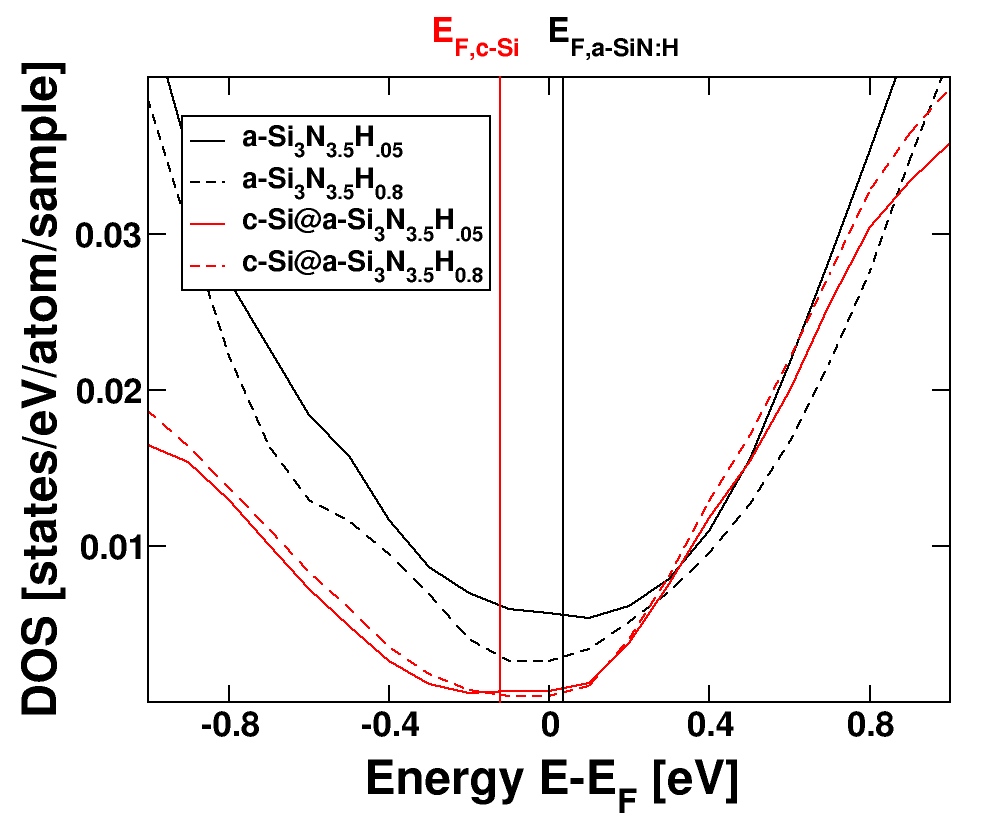}
    \end{center}
    \caption{
(color online) Local electronic density of states evaluated for the c\mbox{-}Si
part and in the a\mbox{-}SiN:H part (red/black lines) of the
c\mbox{-}Si/a\mbox{-}Si$_3$N$_{3.5}$:H interfaces with $1$ and $11$~atm\% H
(solid/dashed lines). The mid gap level of  bulk c\mbox{-}Si and the Fermi
level of a\mbox{-}Si$_3$N$_{3.5}$H$_{0.8}$ are shown by vertical lines in the
corresponding colors.
    }
    \label{fig:dosCsiAsin}
\end{figure}
%-------------------------------------------------------------------------

\subsection{Electronic states in the band gap}

% Energy distribution of gap states

To obtain a statistically meaningful result, we now explore the electronic
density of states evaluated close to the interface and averaged over all 90
samples  (Fig.~\ref{fig:dosCsiAsin}). On the c\mbox{-}Si side, we observe a
slight band tailing introduced by local disorder and distorted Si bonds (red
lines). However, the majority of the gap states are caused by undercoordinated
Si atoms at the a\mbox{-}SiN:H side of the interface (black lines).  The most
important observation is that the minimum of the DOS of c\mbox{-}Si is shifted
towards lower energy values (left) in comparison to the minimum of the DOS of
a\mbox{-}SiN:H.  Likewise, the mid gap energy level of bulk c\mbox{-}Si (red
vertical line)--- evaluated by taking the mean value of the valence band
maximum  and conduction band minimum of bulk Si  --- lies about $0.16$~eV below
the average Fermi level of bulk a\mbox{-}SiN:H (black vertical line).  Impaired
by the issue of too small DFT band gaps, we believe that we observe only the
lower bound of this offset here.  Note that we have again taken the core levels
of Si and N as reference for the alignment, and the general observations apply
independently of the H concentration. As for our previous studies on bulk
a\mbox{-}SiN:H \cite{Hintzsche2012,Hintzsche2013}, H mainly reduces the number
of defects, although here it also slightly reduces the misalignment of the
Fermi levels from $0.24$~eV to $0.16$~eV.

% Creation of positve K+ defects at the interface

\begin{figure}
    \begin{center}
        \includegraphics[trim= 0.0cm 0.75cm 0.0cm 0.0cm, width=8.5cm]{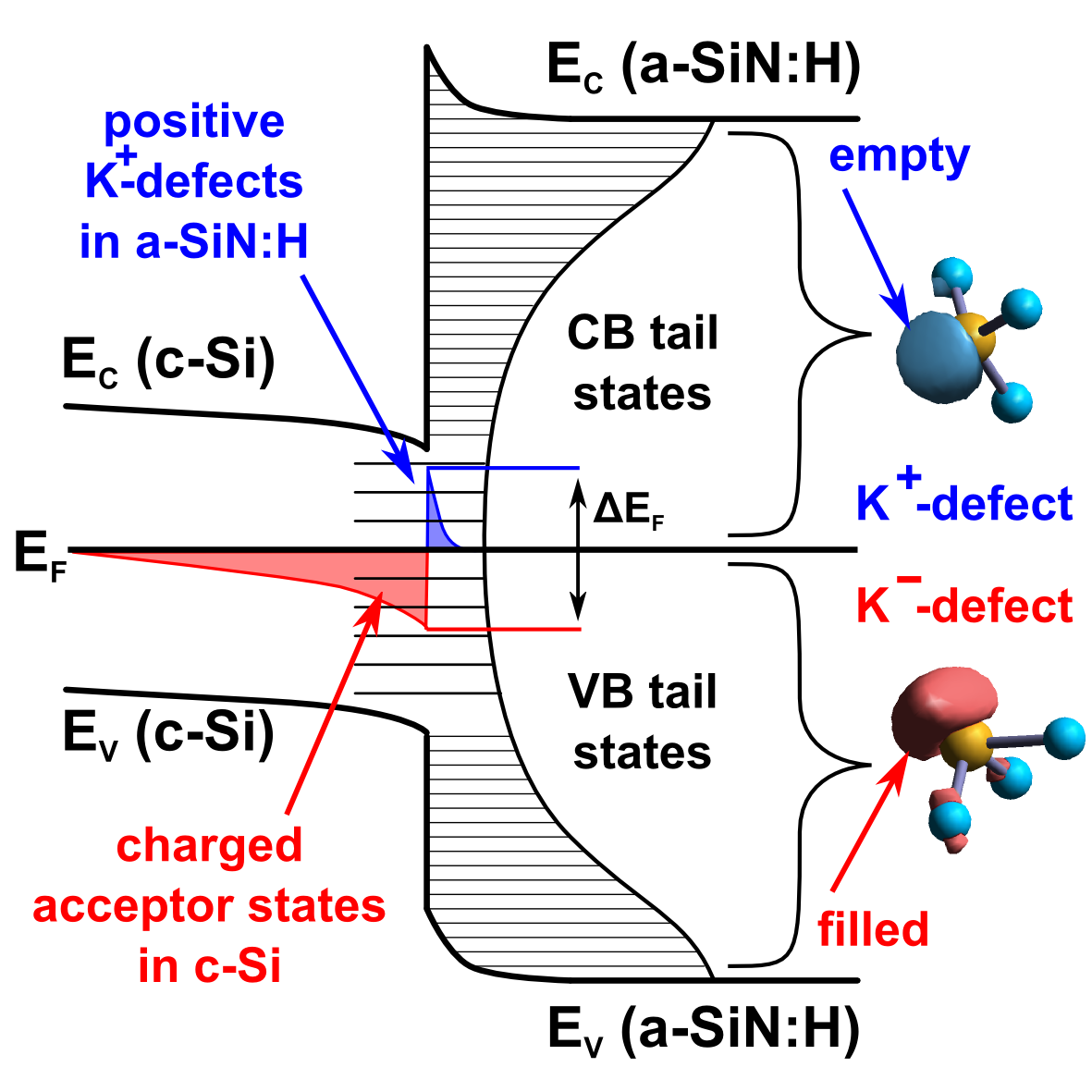}
    \end{center}
    \caption{
(color online) Illustration of the predicted band bending at the
c\mbox{-}Si/a\mbox{-}SiN:H interface. Before bringing both subsystems in
contact, the mid gap level of c\mbox{-}Si lies below the Fermi level of
a\mbox{-}SiN:H (offset between red and blue line at the interface). In
a\mbox{-}SiN:H, nearly all defect levels below and above the Fermi level are
doubly occupied K$^-$ defects and unoccupied K$^+$ defects, respectively. Only
a few singly occupied K$^0$ defects are located at the Fermi level. To align
the Fermi levels, K$^-$ and K$^0$ defects donate charge to shallow acceptor
levels in c\mbox{-}Si (red area). They are thereby converted to K$^0$ and K$^+$
defects (blue area).
}
    \label{fig:ChargeTransfer}
\end{figure}

The offset of the mid gap level of Si and the Fermi level of a\mbox{-}SiN:H has
important consequences. In a macroscopic sample, the Fermi levels must align
and, in a real device, the alignment would occur via charge transfer from
donors, here doubly occupied K$^-$ defects in a\mbox{-}SiN:H, to acceptor
levels in c\mbox{-}Si as illustrated in Fig.~\ref{fig:ChargeTransfer}. In our
microscopic realizations this is not strictly so, since the Fermi level of the
c\mbox{-}Si part can be located anywhere between the conduction and valence
band: in the absence of acceptors and donors, the c\mbox{-}Si part does not pin
the Fermi level mid gap. Only if an occupied K$^-$ defect is located above the
conduction band of c\mbox{-}Si, a partial charge transfer to the c\mbox{-}Si
part is observed, like in sample C. This partial charge transfer shifts the
c\mbox{-}Si states upwards and the a\mbox{-}SiN:H states downwards.  This line
of thought indicates that it is not quite straightforward to determine highly
accurate conduction band and valence band offsets between amorphous model
structures and crystalline samples by using small model structures, even though
we average over many microscopic models. That the mid gap level  of c\mbox{-}Si
lies well below the Fermi level of a\mbox{-}SiN:H is, however, a robust feature
corroborated by three independent observations: (i) the observation of
microscopic models with charge transfer to c\mbox{-}Si, (ii) the visual
misalignment of the minima in the electronic DOSs, and (iii) the misalignment
of the calculated c\mbox{-}Si mid gap level and the Fermi level of
a\mbox{-}SiN:H.

%-------------------------------------------------------------------------
%-------------------------------------------------------------------------
\section{Conclusions}
%  Conclusions

In summary, we conclude that the number of undercoordinated Si atoms (K
defects) is significantly increased at c\mbox{-}Si/a\mbox{-}SiN:H interfaces
compared to bulk a\mbox{-}SiN:H (compare Fig.~\ref{fig:NrDefectProfile}(b)).
This is a result of the structural mismatch between  c\mbox{-}Si and
a\mbox{-}SiN:H. The number of coordination defects and related electronic
defect states is reduced by about a factor $2$ in our simulations when H is
present. Hence, hydrogen clearly cures geometrical defects at the interface,
similar to bulk a\mbox{-}SiN:H. We predict a positive, fixed charge at
c\mbox{-}Si/a\mbox{-}SiN interfaces related to a surplus of K$^+$ defects on
the a\mbox{-}SiN:H side screened by negatively charged acceptor states on the
c\mbox{-}Si side (Fig.~\ref{fig:ChargeTransfer}). In principle, this surplus is
related to the mid gap  level of bulk c\mbox{-}Si lying at least $0.16$~eV
below the Fermi level of bulk a\mbox{-}SiN:H. Consequently, electrons must be
transfered from originally doubly occupied  K$^-$  defects on the
a\mbox{-}SiN:H side, to acceptor levels on the c\mbox{-}Si side, to align the
Fermi levels in the heterostructure. Solving the Poisson equation for this
standard device problem results in a band bending at the interface (indicated
by black lines in Fig.~\ref{fig:ChargeTransfer}). Clearly such a band bending
will attract electrons to the interface aiding the electron removal by metallic
leads, which are etched into the a\mbox{-}SiN:H. Holes, on the other hand, are
repelled making an effective separation of both carriers possible. Our present
study also indicates that the number of K defects and the amount of positive
charge at the interface are interrelated quantities.  If the defect density is
very low, there will be simply no defects that can donate charge to the
c\mbox{-}Si part. This will reduce the positive interface charge and
consequently the carrier separation. If the defect density is high, the fixed
charge density will be high, but more recombination centers will be present as
well. These results are exactly in line with recent experimental studies
\cite{Lamers2012,Lamers2013-2}. Based on our present results, we suppose that
it will be hard to achieve a negative, fixed charge by using the standard
a\mbox{-}SiN:H passivation. The mid gap level of c\mbox{-}Si is well below the
Fermi level of a\mbox{-}SiN:H, and only strong doping might change this.
Moreover, it is impossible to reduce the density of carrier traps (K$^-$,
K$^0$, and K$^+$ defects) without changing the field effect passivation of
fixed charges, as the species responsible for the fixed charges and K defects
are identical.

\section*{Acknowledgement}

This work is part of the HiperSol (high performance solar cells) project funded
by the European commission grant no. MMP3-SL-2009-228513. Supercomputing time
on the Vienna scientific cluster (VSC) is gratefully acknowledged.

\bibliography{References}

\end{document}